\documentclass[12pt]{iopart}
\usepackage{amssymb,dsfont}

\def\pois#1#2{\left\{ {#1},{#2} \right\}}

\def\endpf{\begin{flushright}$\square$\end{flushright}}
\def\al{\alpha}
\def\be{\beta}
\def\ga{\gamma}

\def\back{\!\!\!\!\!\!\!\!\!\!\!\!\!\!\!\!\!}
\def\part#1#2{\frac{\partial {#1}}{\partial {#2}}} 
\def\la{\lambda}

\newtheorem{theorem}{Theorem}

\begin{document}

\title{Generalization of the linear $r-$matrix formulation through Loop coproducts}
\author{Fabio Musso}
\address{Departamento de F\'isica, Universidad de Burgos,
E-09001 Burgos, Spain}
\ead{fmusso@ubu.es}
\begin{abstract}
A new method for the construction of classical integrable systems, that we call loop coproduct formulation, is presented. We show 
that the linear $r-$matrix formulation, the Sklyanin algebras and the reflection algebras can be obtained as particular subcases of this 
framework. We comment on the possible generalizations of the $r-$matrix formalism introduced through this approach.  
\end{abstract}
\section{Introduction}

One of the main tools for the construction and study of classical integrable Hamiltonian systems is the Lax formalism \cite{Lax}. In a noteworthy paper \cite{BV}, Babelon and Viallet showed that any integrable Hamiltonian system with a finite number of degrees of freedom admits a Lax pair, that is 
two $N \times N$ matrices $L(\la),M(\la)$, whose entries are functions on the phase space ($\la$ being the spectral parameter\footnote{Actually, in the original paper \cite{BV}, the spectral parameter does not appear at all, being completely unessential. However, since in many cases the natural Lax pair formulation involves a spectral parameter, we decided to present the results of \cite{BV} in this case.}) and such that the Hamiltonian evolution of the matrix $L(\la)$ is of the form
\begin{equation}
\frac{dL(\la)}{dt}=[L(\la),M(\la)]. \label{Lax}
\end{equation}
Equation (\ref{Lax}) implies that the spectral invariants of the Lax matrix $L(\la)$ are conserved quantities under the Hamiltonian evolution. 
On the other hand Liouville integrability requires the conserved quantities to be in involution among them. In \cite{BV}, Babelon and Viallet
proved that the involution property for the spectral invariants of $L(\la)$ is equivalent to the existence of a linear $r-$matrix formulation
\begin{equation}
\{ L(\la) \otimes \mathds{1}, \mathds{1} \otimes L(\mu) \}=[r_{12}(\la,\mu), L(\la) \otimes \mathds{1}]-[r_{21}(\mu,\la),\mathds{1} \otimes L(\mu)  ], \label{BV}
\end{equation}   
where $L(\la)$ is an $N \times N$ matrix, $r_{12}(\la,\mu)$ is an $N^2 \times N^2$ matrix, $\mathds{1}$ is the $N \times N$ identity matrix and 
$$
r_{21}(\la,\mu)=\Pi \, r_{12}(\la,\mu) \Pi, 
$$
with $\Pi$ being the permutation operator $\Pi (x \otimes y) = y \otimes x, \ \forall x,y \in \mathds{C}^N$. Moreover, they showed that if the Lax matrix
satisfies the equation (\ref{BV}), then the equation of motions associated with any of the spectral invariants of $L(\la)$ admits a Lax representation 
of the form (\ref{Lax}).  

In this paper we present a generalization (in a sense that we will make more precise later) of the linear $r-$matrix formalism that we call 
loop coproduct formulation. A preliminary version of this approach has been presented in \cite{io}, as a generalization of both the Gaudin algebras
(see \cite{Gaudins} and references therein) and the coproduct method \cite{Balrag}, (see \cite{Vulpi} for a recent review on the subject). 
Here we present a more general formulation encompassing
the linear $r-$matrix formulation \cite{BV}, the Sklyanin algebras \cite{Faddeev} and also the reflection equation algebra \cite{Sk}. While all these cases emerge as examples of the loop coproduct formulation applied to the Lie--Poisson algebra ${\cal{F}}(gl(N)^*)$, the loop coproduct
formalism is defined for arbitrary Poisson algebras. We think that the implementation of this approach to non-linear Poisson algebras
could lead to new examples of integrable systems and represents an interesting open problem in the field of classical finite-dimensional integrable systems.
 
\section{Loop coproduct formulation} \label{loopcoproduct}

In this section we present our main theorem, which generalize the result presented in \cite{io}.
Let us consider a generic Poisson algebra $A$ of dimension $M$ with $r$ Casimir functions ${\cal{C}}_j, \ j=1,\dots,r$ and let us denote with $\{y^\alpha\}_{\alpha=1}^M$ a set of generators for $A$ with Poisson brackets:
$$
\{ y^\al, y^\be \}_A=F^{\al \be}(\vec{y}) \qquad \vec{y}=(y^1,\dots,y^M). 
$$
Let $B$ be another Poisson algebra and let us denote with $\vec{Z}$ a set of its generators. 
\begin{theorem}{\label{main}}
Let us consider a set of $m$ maps depending on a parameter $\la$
\begin{equation}
\Delta^{(k)}_{\lambda}: A \to B, \quad k=1, \dots, m \label{lcop}
\end{equation}
and such that the images of the $A$ generators satisfy the following Poisson brackets in $B$:
\begin{eqnarray}
&& \fl \{ \Delta_{\la}^{(i)}(y^\al), \Delta_{\mu}^{(k)}(y^\beta) \}_B=f^\beta_\gamma(i,k,\la,\mu,\vec{Z}) F^{\al \ga} (\Delta_{\la}^{(i)}(\vec{y})) \quad k>i \label{p1}\\
&& \fl \{ \Delta_{\la}^{(k)}(y^\al), \Delta_{\mu}^{(k)}(y^\be) \}_B=g^\beta_\ga(k,\la,\mu,\vec{Z}) F^{\al \ga} (\Delta_{\la}^{(k)}(\vec{y}))+ h^\al_\ga(k,\la,\mu,\vec{Z}) F^{\ga \be} (\Delta_{\mu}^{(k)}(\vec{y}))\label{p2}
\end{eqnarray}
for certain functions $f^\beta_\ga(i,k,\la,\mu,\vec{Z}),g^\beta_\ga(k,\la,\mu,\vec{Z}),h^\al_\ga(k,\la,\mu,\vec{Z})$.

If the map $\Delta_{\la}^{(i)}$ is defined on any smooth function of the generators $G \in A$ as: 
\begin{equation}
\Delta^{(k)}_{\lambda}(G)(y^1,\dots,y^M)) = G(\Delta^{(k)}_{\lambda}(y^1),\dots,\Delta^{(k)}_{\lambda}(y^M)), \label{extension}
\end{equation}
then:
\begin{eqnarray}
 \{ \Delta^{(i)}_{\lambda}({\cal{C}}_j) ,  \Delta^{(k)}_\mu (y^\be) \}_B&=&0 \quad k > i  \label{lambda1} \\
 \{ \Delta^{(i)}_{\lambda}({\cal{C}}_j) ,  \Delta^{(k)}_{\mu}({\cal{C}}_l) \}_B&=&0.  \label{lambda2}
\end{eqnarray} 
\end{theorem}  

\noindent {\bf Proof:} 

\noindent Let us prove first the equation (\ref{lambda1}). We fix an arbitrary Casimir ${\cal{C}}_j$ of $A$.
Since ${\cal{C}}_j$ is a Casimir function, for any $\beta$ we must have:
\begin{equation}
\pois{{\cal{C}}_j}{y^\be}_A=\sum_{\al=1}^M \part{{\cal{C}}_j}{y^\al} \pois{y^\al}{y^\be}_A=\sum_{\al=1}^M \part{{\cal{C}}_j}{y^\al} 
F^{\al \be}(\vec{y}) =0.  \label{classic} 
\end{equation}
Now we use this equation together with (\ref{p1}) to prove equation (\ref{lambda1}):
\begin{eqnarray*}
&& \back \{ \Delta^{(i)}_{\lambda}({\cal{C}}_j),  \Delta^{(k)}_\mu (y^\be) \}_B=\sum_{\al=1}^{M}  \part{{\cal{C}}_j \left(\Delta^{(i)}_{\lambda}(\vec{y}) \right)}{\Delta^{(i)}_\la (y^\al)} \,
\pois{\Delta^{(i)}_\la (y^\al)}{\Delta^{(k)}_\mu (y^\be) }_B=\\  
&&\back =\sum_{\ga=1}^M f^\beta_\gamma(i,k,\la,\mu,\vec{Z}) \sum_{\al=1}^{M} \part{{\cal{C}}_j \left(\Delta^{(i)}_{\lambda}(\vec{y}) \right)}{\Delta^{(i)}_\la (y^\al)} \,  F^{\al \ga} (\Delta_{\la}^{(i)}(\vec{y}))=0   \qquad k>i.
\end{eqnarray*}
From equation 
$$
 \{ \Delta^{(i)}_{\lambda}({\cal{C}}_j) ,  \Delta^{(k)}_\mu(y^\al) \}_B=0 \qquad k > i,
$$
it follows that 
$$
\{\Delta^{(i)}_{\lambda}({\cal{C}}_j),  \Delta^{(k)}_{\mu}({\cal{C}}_l) \}_B=0 \qquad i \neq k,
$$
so that only the case $k=i$ remains to be proven. In this case:
\begin{displaymath}
\fl \{ \Delta^{(i)}_{\lambda}({\cal{C}}_j) ,  \Delta^{(i)}_{\mu}({\cal{C}}_l) \}_B= \sum_{\al,\be=1}^M \part{{\cal{C}}_j \left(\Delta^{(i)}_{\lambda}(\vec{y}) \right)}{\Delta^{(i)}_\la (y^\al)}   \part{{\cal{C}}_l \left(\Delta^{(i)}_{\mu}(\vec{y}) \right)}{\Delta^{(i)}_\mu (y^\be)} \,\pois{\Delta^{(i)}_\la (y^\al)}{\Delta^{(i)}_\mu (y^\be) }_B.
\end{displaymath} 
By using equation (\ref{p2}) and (\ref{classic}), we have:
\begin{eqnarray*}
&& \fl \{ \Delta^{(i)}_{\lambda}({\cal{C}}_j) ,  \Delta^{(i)}_{\mu}({\cal{C}}_l) \}_B=\\
&& \fl = \sum_{\be,\ga} \part{{\cal{C}}_l \left(\Delta^{(i)}_{\mu}(\vec{y}) \right)}{\Delta^{(i)}_\mu (y^\be)} 
g^\beta_\ga(k,\la,\mu,\vec{Z}) \left( \sum_{\al=1}^{M}  \part{{\cal{C}}_j \left( \Delta^{(i)}_{\lambda}(\vec{y})\right)}{\Delta^{(i)}_\la (y^\al)} F^{\al \ga}( \Delta_{\la}^{(i)}(\vec{y})) \right) + \\
&& \fl + \sum_{\al,\ga} \part{{\cal{C}}_j \left(\Delta^{(i)}_{\la}(\vec{y}) \right)}{\Delta^{(i)}_\la (y^\al)} 
h^\al_\ga(k,\la,\mu,\vec{Z}) \left( \sum_{\be=1}^{M}  \part{{\cal{C}}_l \left( \Delta^{(i)}_{\mu}(\vec{y})\right)}{\Delta^{(i)}_\mu (y^\be)} F^{\ga \be}( \Delta_{\mu}^{(i)}(\vec{y})) \right)=0
\end{eqnarray*} 
since the terms in parentheses vanish, which completes the proof. 
\endpf
We will call the maps (\ref{lcop}) satisfying equations (\ref{p1}) and (\ref{p2}) ``loop coproducts'' and we will say that the integrable systems
that can be obtained from equations (\ref{lambda1}),(\ref{lambda2}) ``admit a loop coproduct formulation''.
 
\section{Particular cases} 
Let us now show that the linear $r-$matrix formulation (\ref{BV}) is a subcase of the loop coproduct formulation.
In order to prove this result, let us take $A$ as the Lie-Poisson algebra ${\cal{F}}(gl(N)^*)$, with the standard set of generators $\{ e_{ij} \}, \ i,j=1,\dots,N$:
\begin{equation}
\{ e_{ij}, e_{kl} \}= \delta_{jk} e_{il} - \delta_{il} e_{kj} \label{gln}.
\end{equation}    
The Poisson algebra $B$ will be the one on which the $r$-matrix formulation (\ref{BV}) is defined. The map 
$\Delta_{\lambda}: A \to B$ 
is defined on the $A$ generators as 
\begin{equation}
\Delta_{\lambda}(e_{ij})=L_{ij}(\la) \label{glnmap}
\end{equation}
and is extended to an arbitrary element of $A$ through equation (\ref{extension}). Let us denote with $E^{ij}$ the canonical $gl(N)$ matrix generators
$$
\left(E^{ij} \right)_{kl}=\delta_{i k} \delta_{j l}
$$
and let us compute the Poisson bracket (\ref{p2}) in this case:
\begin{eqnarray}
&& \fl \{ \Delta_{\lambda}(e_{ij}), \Delta_{\lambda}(e_{kl}) \}_B= \{ L_{ij}(\la), L_{kl}(\mu) \}=  
{\rm Tr} \left( \{ L(\la) \otimes \mathds{1}, \mathds{1} \otimes L(\mu) \} E^{ji} \otimes E^{lk} \right)= \nonumber\\
&& \fl = {\rm Tr} \left( \left\{  [r_{12}(\la,\mu), L(\la) \otimes \mathds{1}]- [r_{21}(\mu,\la),\mathds{1} \otimes L(\mu)  ]
 \right\}E^{ji} \otimes E^{lk} \right)= \nonumber\\
 && \fl = {\rm Tr} \left( [r_{ab,cd}(\la,\mu) (E^{ab} \otimes E^{cd}), L_{ef}(\la) (E^{ef} \otimes \mathds{1})] E^{ji} \otimes E^{lk} \right)- \nonumber \\
 && \fl - {\rm Tr} \left( [r_{ab,cd}(\mu,\la) (E^{cd} \otimes E^{ab}), L_{ef}(\mu) (\mathds{1} \otimes E^{ef}  )] E^{ji} \otimes E^{lk} \right)= \nonumber \\
 && \fl =\sum_{a} r_{ia,kl}(\la,\mu) L_{aj}(\la) - r_{aj,kl}(\la,\mu) L_{ia}(\la)-r_{ka,ij}(\mu,\la) L_{al}(\mu)+r_{al,ij}(\mu,\la) L_{ka}(\mu). \label{last}   
\end{eqnarray}
On the other hand, the right hand side of the equation (\ref{p2}), with the choices (\ref{gln}) and (\ref{glnmap}), is given by:  
\begin{equation}
\fl \sum_{a,b} g^{kl}_{ab} (k,\la,\mu,\vec{Z}) \left( \delta_{ja} L_{ib} (\la) - \delta_{ib} L_{aj} (\la) \right) + h^{ij}_{ab} (k,\la,\mu,\vec{Z}) 
\left( \delta_{bk} L_{al} (\mu) - \delta_{al} L_{kb} (\mu) \right) \label{rhs}
\end{equation}
and (\ref{rhs}) coincides with the last line of (\ref{last}) when 
\begin{equation}
g^{kl}_{ab} (k,\la,\mu,\vec{Z})= -r_{ba,kl}(\la,\mu) \qquad h^{ij}_{ab}(k,\la,\mu,\vec{Z})=-r_{ba,ij} (\mu,\la). \label{choice}
\end{equation}
Let us stress that the $r-$matrix in equation (\ref{choice}) can be of dynamical type, since the functions $g^{kl}_{ab}$ and $h^{ij}_{ab}$ are 
arbitrary functions of the $B$ manifold coordinates. We also point out that the images of the ${\cal{F}}(gl(N)^*)$ Casimirs ${\cal{C}}_i$, 
$i=1,\dots,N$ under the loop coproduct (\ref{glnmap}) coincide with the spectral invariants of the Lax matrix $L(\la)$:
$$
\Delta_{\la}({\cal{C}}_i)=\Tr(L(\la)^i), \qquad i=1,\dots,N.
$$
When the $r$-matrix is non-dynamical and unitary ($r_{12}(\la,\mu)=r(\la-\mu)=-r_{21}(\mu,\la)$), then 
equation (\ref{BV}) reduces to 
$$
\{ L(\la) \otimes \mathds{1}, \mathds{1} \otimes L(\mu) \}=[r(\la-\mu), L(\la) \otimes \mathds{1} + \mathds{1} \otimes L(\mu)  ] 
$$
and defines the so called Gaudin algebras (see \cite{Gaudins} and references therein). This case has been extensively considered, in the framework of the loop coproduct formulation, 
in \cite{io}. 
For such $r-$matrices, one can also define the Sklyanin algebra \cite{Faddeev}
\begin{equation}
\{ L(\la) \otimes \mathds{1}, \mathds{1} \otimes L(\mu) \}=[r(\la-\mu), L(\la) \otimes  L(\mu)  ], \label{Sklyanin}
\end{equation}  
which is quadratic in the Lax matrix entries. The Sklyanin algebras are also a subcase of the loop coproduct formulation with the same loop coproduct as 
defined by (\ref{glnmap}) and the particular choice
\begin{eqnarray*}
g^{kl}_{ab}(k,\la,\mu,\vec{Z})&=& -\frac12 \sum_c \left( r_{ba,kc}(\la-\mu) L_{cl}(\mu)+ r_{ba,cl}(\la-\mu) L_{kc}(\mu) \right), \\
h^{ij}_{ab}(k,\la,\mu,\vec{Z})&=&  \frac12 \sum_c \left( r_{ic,ba}(\la-\mu) L_{cj}(\la) + r_{cj,ba}(\la-\mu) L_{ic}(\la) \right). 
\end{eqnarray*}
This is a plain consequence of the fact that equation (\ref{Sklyanin}) can be reformulated into the form (\ref{BV}) by introducing the dynamical $r-$matrix
$$
r_{12}(\la,\mu)=\frac12 \left( r(\la-\mu) (\mathds{1} \otimes L(\mu)) + (\mathds{1} \otimes L(\mu)) r(\la-\mu) \right). 
$$
Another example of a quadratic Poisson algebra associated with a unitary non-dynamical $r-$matrix is the reflection algebra \cite{Sk}:
\begin{eqnarray*}
&& \fl \{ L(\la) \otimes \mathds{1}, \mathds{1} \otimes L(\mu) \}=[r(\la-\mu), L(\la) \otimes  L(\mu)  ]+  \\
&& \fl + ( L(\la) \otimes \mathds{1} ) r(\la+\mu) (\mathds{1} \otimes L(\mu))-(\mathds{1} \otimes L(\mu)) r(\la+\mu) ( L(\la) \otimes \mathds{1} ). 
\end{eqnarray*}
It is immediate to check that this algebra is also a subcase of the loop coproduct formulation with the same loop coproduct as 
defined by (\ref{glnmap}) provided that
\begin{eqnarray*}
g^{kl}_{ab}(k,\la,\mu,\vec{Z})&= \sum_c \left( r_{ba,kc}(\la+\mu) L_{bl}(\mu)- r_{ba,cl}(\la-\mu) L_{kc}(\mu) \right), \\
h^{ij}_{ab}(k,\la,\mu,\vec{Z})&= \sum_c \left( r_{ic,ba}(\la-\mu) L_{cj}(\la)+ r_{ic,ba}(\la+\mu) L_{cj}(\la) \right). 
\end{eqnarray*}

\section{Discussion and perspectives}
While it is true that any integrable system admits, at least in a neighborhood of generic points of the phase space, an $r-$matrix formulation \cite{BV} (so that, as a consequence, we can state that any integrable system admits a loop coproduct formulation), the proof of the existence of the Lax matrix $L(\la)$ relies on the knowledge of the action-angle coordinates, so that, in general, finding it could be not easier than solving 
explicitly the associated dynamical system.
On the other hand, for some Hamiltonian systems, the $r-$matrix approach turns out to be a precious instrument to establish the integrability of the system, and such systems can be considered to possess a ``natural'' $r-$matrix formulation. In the same sense there will be Hamiltonian systems that admits a ``natural'' loop coproduct formulation. We claim that the class of systems of the latter kind is potentially much larger than those of the former one. Indeed, as we have shown, the linear $r-$matrix formulation can be seen as a subcase of the loop-coproduct formulation when
\begin{enumerate}
\item there is a unique loop-coproduct $\Delta_{\lambda}: A \to B$, 
\item $A$ is the Lie--Poisson algebra ${\cal{F}}(gl(N)^*)$.    
\end{enumerate}
Hence, we can generalize the linear $r-$matrix approach by looking for Hamiltonian systems admitting a natural loop coproduct formulation with
\begin{enumerate}
\item $k>1$ loop coproduct maps $\Delta^{(i)}_{\lambda}: A \to B, \quad i=1,\dots,k$, 
\item $A$ being a non-linear Poisson algebra.    
\end{enumerate}
Examples of systems of the first kind emerge naturally in the so called coproduct method \cite{Balrag}, \cite{Vulpi}. Namely, in \cite{io} it has been proven that all the integrable systems coming from the coproduct method admit a natural loop coproduct formulation, so that one can define $k$ loop 
coproduct maps (\ref{lcop}) satisfying (\ref{p1}) and (\ref{p2}) and the integrability of the system can be deduced from equations (\ref{lambda1}) and (\ref{lambda2}).
Typically, in such systems, one can associate with each map a linear $r-$matrix that assures that 
\begin{equation}
\{\Delta^{(i)}_{\lambda}({\cal{C}}_j),  \Delta^{(i)}_{\mu}({\cal{C}}_l) \}_B=0 \label{equal}
\end{equation}
but notice that the following relations also hold:
\begin{equation}
\{\Delta^{(i)}_{\lambda}({\cal{C}}_j),  \Delta^{(k)}_{\mu}({\cal{C}}_l) \}_B=0 \qquad k \neq i. \label{unequal}
\end{equation}
We stress that finding a unique $r-$matrix formulation of the form (\ref{BV}) accounting for both (\ref{equal}) and (\ref{unequal}) seems not to be an easy task. 
While this example shows that the loop coproduct formulation should be indeed regarded as more general of the $r-$matrix one (in the sense previously explained), the 
second kind of generalization is, in our opinion, the really interesting one. 
Therefore,
we suggest that finding a loop coproduct
formulation for a non-linear Poisson algebra $A$ could lead to the construction of new classes of integrable systems and we propose it as an open research line in the field of classical, integrable and 
finite-dimensional Hamiltonian systems.

\section*{Acknowledgments}
The author thanks A. Ballesteros, F.J. Herranz, M. Petrera and O. Ragnisco for useful discussions. This work was partially supported by the Spanish MICINN under grant MTM2007-67389 (with EU-FEDER support), by Junta de Castilla y Le\'on (Project GR224) and by INFN-CICyT.

\end{document}